\documentstyle[aps,amsfonts]{revtex}
\def\Slash#1{{#1\llap{/}}}
\def\tr#1{{\text{tr}\left(#1\right)}}
\def\Re{\text{Re}\,}
\def\one{\mbox{\bf 1}}
\def\zero{\mbox{\bf 0}}
\def\Cl{{\cal C \it l}}
\def\Bar#1{{{#1}^*}}
\def\Sphf{A}
\def\adj#1{\overline{#1}}
\def\T#1{{{#1}^T}}
\def\opp{{\text{opp}}}
\def\vm{\mathbin{\lower.25ex\hbox{$\scriptscriptstyle\vee$}}}
\def\cpr_#1{\mathbin{%
  \mathchoice
    {\displaystyle{\mathop{\circ}_{\raise.7ex\hbox{$\scriptstyle#1$}}}}
    {\displaystyle{\mathop{\textstyle\circ}_%
       {\textstyle\raise.7ex\hbox{$\scriptstyle#1$}}}}
    {\displaystyle{\mathop{\scriptstyle\circ}_%
       {\scriptstyle\raise.7ex\hbox{$\scriptscriptstyle#1$}}}}
    {\displaystyle{\mathop{\scriptscriptstyle\circ}_%
       {\scriptscriptstyle\raise.7ex\hbox{$\scriptscriptstyle#1$}}}}}}
\def\JO#1{{\cal #1}}
\def\jo{\circ}
\def\twobytwo#1{\left(\begin{array}{cc}#1\end{array}\right)}
\def\Z{{\cal Z}}
\def\fr{{*}}
\def\unit#1{{\breve{#1}}}
\def\R{{\Bbb R}}
\def\C{{\Bbb C}}
\def\O{{\Bbb O}}
\def\Zeta{\text{Z}}
\let\isom = \approx
\let\implies = \Longrightarrow
\edef\SC{\ifpreprintsty ;\\ \else ;\ \fi}
\def\acsz{\arraycolsep 0pt}
\begin{document}
%
\title{The general classical solution of the superparticle}
\author{J\"org Schray}
\address{Department of Physics, Oregon State University,
		Corvallis, OR  97331, USA \\
{\tt schrayj{\rm @}physics.orst.edu}
}
\date{May 12, 1994}
\maketitle
\begin{abstract}
The theory of vectors and spinors in 9+1 dimensional spacetime is introduced
in a completely octonionic formalism based on an octonionic representation of
the Clifford algebra $\Cl(9,1)$.
The general solution of the classical equations of motion of the CBS
superparticle is given to all orders of the Grassmann hierarchy.  A spinor and
a vector are combined into a $3 \times 3$ Grassmann, octonionic, Jordan matrix
in order to construct a superspace variable to describe the superparticle.
The combined Lorentz and supersymmetry transformations of the fermionic and
bosonic variables are expressed in terms of Jordan products.
\end{abstract}


\section{Introduction}

The relationship between the division algebras and the existence of
supersymmetric theories has been observed before \cite{supersymmetry and
octonions,Gursey 5by5,Chung}, especially in the context of string theory.
In particular, the division algebras have been used to solve the classical
equations of motion for the superparticle and the superstring
\cite{superstring and superparticle}.
However, because of the non-associativity of the octonions, there have been
difficulties in the the case of this division algebra of highest dimension.
For example, the Lorentz invariance of the formalism was unclear.
Since the CBS superparticle \cite{CBS} is an ideal testing ground to introduce
techniques using division algebras and explore supersymmetry, it has attracted
some attention; see I.~Oda {\it et al.\/}\cite{Oda} and H.~Tachibana \&
K.~Imaeda\cite{Tachibana}.
The connection of supersymmetric theories to the division algebras can also be
made in terms of Jordan algebras as in \cite{Jordan} and especially \cite{Foot
and Joshi}.

This article carries on these previous attempts to cast the description of the
superparticle in a form
that clearly displays its symmetries; however, a more transparent and powerful
octonionic formalism is used.  We go beyond a mere rewriting of vector and
spinor variables in terms of octonionic expressions, with supersymmetry and
Lorentz transformations acting differently on these variables.  We succeed in
introducing a unified superspace variable as a Jordan matrix, which includes
both fermionic and bosonic variables.  Both the supersymmetry transformations
and the general solution are expressed in terms of Jordan matrices involving
both kinds of variables in this unified way.

There are many other approaches, not involving division algebras, which
investigate supersymmetry in the superparticle, see for example \cite{others}.

This article is organized as follows.  Section \ref{sec:octonions} introduces
the octonionic formalism for vectors and spinors and their Lorentz
transformations in 9+1 dimensions.  (We deal exclusively with the
9+1-dimensional case. The analogues in 5+1, 3+1, and 2+1 dimensions can be
found easily.)  A subsection using the octonionic analogue of the Fierz-matrix
\cite{Fierz} derives what we call the 3-$\Psi$'s rule, an identity that is
needed for the Green-Schwarz superstring to be supersymmetric
\cite{Green-Schwarz}.  A note on the notion of octonionic dotted and undotted
spinors concludes this introductory section.  Section \ref{sec:superparticle}
derives the general classical solution of the equations of motion for the CBS
superparticle.  Section \ref{sec:Jordan} develops the Jordan matrix formalism
combining bosonic and fermionic variables into one object.  Lorentz and
supersymmetry transformations and the superparticle action are expressed in
this way.

\section{Octonionic spinors and the 3-$\Psi$'s rule}\label{sec:octonions}

\subsection{Octonionic spinors}

Octonionic spinors are based on an octonionic representation of a Clifford
algebra.  The non-associativity of the octonions raises obstacles which can be
removed with care.  A rigorous treatment and resolution of this issue can be
found in \cite{Clifford}, which also contains an introduction to octonions.
Only general properties of octonions independent of a specific multiplication
table will be used here.  However, because we make frequent use of a variety
of octonionic identities, the reader may find more information on octonions
helpful; see \cite{Lorentz,octonions}.

The full Clifford algebra $\Cl(9,1)$ in 9+1 dimensions has a real, faithful,
irreducible, Weyl representation in terms of $32\times32$-matrices.  (As a
reference for the general topic of Clifford algebras see \cite{general
Clifford,Clifford}.)  An octonionic Majorana-Weyl representation is given in
terms of $4\times4$-matrices:
\begin{equation}
\gamma_\mu
=\left(
\begin{array}{cc}
    \zero & \Gamma_\mu \\
    \tilde \Gamma_\mu & \zero
\end{array}
\right),
\label{eq:9-1rep}
\end{equation}
where
\begin{equation}\acsz\begin{array}{rl}
\Gamma_0 ={}&-\tilde \Gamma_0 = \one
= \left( \begin{array}{cc} 1 & 0 \\ 0 & 1 \end{array}\right),\\
\noalign{\medskip}
\Gamma_j ={}&\tilde \Gamma_j
= \left( \begin{array}{cc} 0 & e_j \\ \Bar{e_j} & 0 \end{array}\right)
\qquad (1\leq j\leq 8),\\ \noalign{\medskip}
\Gamma_9 ={}&\tilde \Gamma_9
= \left( \begin{array}{cc} 1 & \phantom{-}0 \\ 0 & -1 \end{array}\right),
\\ \noalign{\medskip}
\gamma_{11} ={}&\gamma_0 \gamma_1 \ldots \gamma_9
= \left( \begin{array}{cc} \one & \phantom{-}\zero \\ \zero & -\one \end{array}
\right).
\end{array}\label{eq:9-1even rep}\end{equation}
(In our conventions an octonion $x$ has real components $x^j \;\; (1\leq j\leq
8)$, i.e., $x=x^je_j$, where $e_j \;\; (1\leq j\leq 8)$ are the octonionic
units and $\Bar{e_j}$ their octonionic conjugates.
The signature of the metric is $-+\ldots+$.)
This representation is understood to act on a column of four octonions, a
spinor, by left multiplication.  This notion is necessary in order to remove
the ambiguity that arises from the fact that octonionic multiplication is not
associative.  The fundamental property $\gamma_\mu \gamma_\nu + \gamma_\nu
\gamma_\mu = 2 g_{\mu\nu}$ remains valid under this interpretation.  (For a
rigorous treatment see \cite{Clifford}.)

A vector with components $x^\mu \;\; (0\leq \mu\leq 9)$ is embedded in the
Clifford algebra via
\begin{equation}
\Slash{x} = x^\mu \gamma_\mu =
\left( \begin{array}{cc} 0 & \bf X \\ \tilde {\bf X} & 0 \end{array} \right),
\end{equation}
where
\begin{equation}
{\bf X} = x^\mu \Gamma_\mu \quad\text{\ and\ }\quad
\tilde {\bf X} = x^\mu \tilde \Gamma_\mu.
\label{eq:boldface}
\end{equation}
(Boldface capitals always denote the $2\times 2$ hermitian matrix associated
as in (\ref{eq:boldface}) with the vector denoted by the same lowercase
letter.)  The inverse of this relationship is
\begin{equation}
x^\mu= {1\over 4}\Re\tr{\Slash{x} \gamma^\mu}
= {1\over 4}\Re\tr{{\bf X}\tilde\Gamma^\mu + \tilde {\bf X}\Gamma^\mu}
= {1\over 2}\Re\tr{{\bf X}\tilde\Gamma^\mu}
= {1\over 2}\Re\tr{\tilde {\bf X}\Gamma^\mu},
\end{equation}
where indices are raised with the metric tensor $g$.
Also note that
\begin{equation}
\Gamma^\mu=\tilde \Gamma_\mu
\end{equation}
and
\begin{equation}
\tilde {\bf X} = {\bf X} - (\tr{\bf X})\,\one,
\label{eq:tilde}
\end{equation}
which implies
\begin{equation}
{\bf X}\tilde {\bf X} = \tilde {\bf X} {\bf X}
= {\bf X}^2 - (\tr{\bf X})\,{\bf X}
= -\text{det}({\bf X})\,\one,
\label{eq:x-xtilde}
\end{equation}
since the characteristic polynomial for a hermitian $2\times 2$-matrix $A$ is
$p_A(z) = z^2 - \tr{A} z + \text{det}(A)$.
The combination (\ref{eq:x-xtilde}) appears in the matrix product
\begin{equation}
\Slash{x}\Slash{x} = x_\mu x^\mu \one_{4 \times 4},
\end{equation}
so that we have
\begin{equation}
x_\mu x^\mu \one = {\bf X}\tilde {\bf X} = -\text{det}({\bf X})\one
\end{equation}
or its polarized form,
\begin{equation}\acsz\begin{array}{rl}
2 x_\mu y^\mu \one ={}& {\bf X}\tilde {\bf Y} + {\bf Y}\tilde {\bf X} \\
\noalign{\smallskip}
={}&\tilde {\bf X} {\bf Y} + \tilde {\bf Y} {\bf X}.
\end{array}\end{equation}
Now our convention for the numbering of the components of an octonion allows
us to simply write
\begin{equation}
{\bf X} =
\left( \begin{array}{cc}
x^+ & x^{\phantom{-}} \\
\Bar{x} & x^-
\end{array} \right) \text{\ and\ }\>
\tilde {\bf X} =
\left( \begin{array}{cc}
-x^- & \phantom{-}x^{\phantom{+}} \\
\phantom{-}\Bar{x} & -x^+ \end{array} \right)
\text{, where\ } x^\pm=x^0 \pm x^9.
\end{equation}

A full spinor $\Psi$ is given by a column of four arbitrary octonions. It can
be decomposed into its positive and negative chiral projections,
\begin{equation} \Psi_\pm := P_\pm \Psi, \end{equation}
via the projection operators
\begin{equation} P_\pm= {1\over 2}(1 \pm \gamma_{11}). \end{equation}
For the chiral projections either the top or the bottom two components vanish.
Depending on the context we will often regard a chiral spinor as just the
column of the two non-vanishing components. We also define the adjoint
spinor:
\begin{equation} \adj{\Psi} := \Psi^\dagger \Sphf.
\label{eq:D-adjoint}\end{equation}
$\Sphf$ is the matrix that intertwines the given representation with the
hermitian conjugate representation:
\begin{equation} \gamma_\mu^\dagger \Sphf = \Sphf \gamma_\mu.\end{equation}
Then $\Sphf$ is given up to a constant by
\begin{equation}
\Sphf = \gamma_0\gamma_{11} =
\left( \begin{array}{cc}
\zero & \one \\ \one & \zero
\end{array}\right).
\end{equation}
($\,{}^\dagger$ denotes matrix transposition composed with octonionic
conjugation.)
The construction of a vector $y$ out of two spinors $\Phi$ and $\Psi$ is done
in the usual way:
\begin{equation}
\acsz\begin{array}{rl}
y_\mu :={}& \Re \left[\adj{\Phi} \gamma_\mu \Psi\right]\\ \noalign{\medskip}
={}&\Re \left[
\begin{array}{cc} ({\Phi_-}^\dagger & {\Phi_+}^\dagger)
\\ \phantom{{\Phi_-}^\dagger}
\end{array}
\left(\begin{array}{cc}
\zero & \Gamma_\mu \\ \tilde \Gamma_\mu & \zero
\end{array}\right)
\left(\begin{array}{c} \Psi_+\\ \Psi_- \end{array}\right)
\right]\\ \noalign{\medskip}
={}& \Re \left[ {\Phi_+}^\dagger \tilde\Gamma_\mu \Psi_+ \right]
+ \Re \left[ {\Phi_-}^\dagger \Gamma_\mu \Psi_- \right].
\end{array}\label{eq:previous}
\end{equation}

So far we have built everything out of real octonions, i.e., the components
$x^j$ of an octonion $x=x^je_j$ were real numbers.  However, in order to
consider anticommuting spinors we need to introduce elements of a Grassmann
algebra.  This notion can be incorporated into the octonionic formalism by
letting the octonionic components take values in a real Grassmann algebra of
arbitrary, possibly infinite, dimension.
(Formally, we extend the octonionic algebra by taking the tensor product with
a Grassmann algebra over a real vector space, e.g., octonionic conjugation
only affects the octonionic part.)
Then the components of the octonions that make up an anticommuting spinor are
odd Grassmannian.  For the previous relation (\ref{eq:previous}) we now
obtain, in addition:
\begin{equation}
\acsz\begin{array}{rl}
y_\mu ={}& \Re \left[\adj{\Phi} \gamma_\mu \Psi\right]\\ \noalign{\medskip}
={}& -\Re \left[\adj{\Psi} \gamma_\mu \Phi\right]\\ \noalign{\medskip}
={}& -\Re \left[ {\Psi_+}^\dagger \tilde\Gamma_\mu \Phi_+\right]
- \Re \left[ {\Psi_-}^\dagger \Gamma_\mu \Phi_- \right].
\end{array}
\end{equation}
The cyclic properties of the trace and the vanishing of the real parts of
graded commutators and associators imply the following identities:
\begin{equation}
\acsz\begin{array}{rl}
y_\mu
={}& -\Re\tr{\Psi \adj{\Phi} \gamma_\mu}\\ \noalign{\medskip}
={}& \Re\tr{\Phi \adj{\Psi} \gamma_\mu}\\ \noalign{\medskip}
={}& -\Re\tr{\Psi_+ {\Phi_+}^\dagger \tilde\Gamma_\mu}
- \Re\tr{\Psi_-{\Phi_-}^\dagger \Gamma_\mu}\\ \noalign{\medskip}
={}& \Re\tr{\Phi_+ {\Psi_+}^\dagger \tilde\Gamma_\mu}
+ \Re\tr{\Phi_-{\Psi_-}^\dagger \Gamma_\mu}\\ \noalign{\medskip}
={}&{1\over 2}\Re\tr{(\Phi_+ {\Psi_+}^\dagger - \Psi_+ {\Phi_+}^\dagger)
\tilde\Gamma_\mu}
+ {1\over 2}\Re\tr{(\Phi_-{\Psi_-}^\dagger - \Psi_-{\Phi_-}^\dagger)
\Gamma_\mu}.
\end{array}\label{eq:ymu}
\end{equation}
The full power of the octonionic formalism becomes evident when we write $y$
in terms of its Clifford representation ${\bf Y}$ and $\tilde{\bf Y}$ without
the use of the Dirac matrices, as follows:
\begin{equation}\acsz\begin{array}{rl}
{\bf Y} ={}&  (\Phi_+ {\Psi_+}^\dagger - \Psi_+ {\Phi_+}^\dagger)
+ \widetilde{(\Phi_-{\Psi_-}^\dagger - \Psi_-{\Phi_-}^\dagger)},\\
\noalign{\medskip}
\tilde{\bf Y} ={}
&\widetilde{(\Phi_+ {\Psi_+}^\dagger - \Psi_+ {\Phi_+}^\dagger)}
+ (\Phi_-{\Psi_-}^\dagger - \Psi_-{\Phi_-}^\dagger).
\end{array}
\label{eq:Fierz1}
\end{equation}
This form of writing ${\bf Y}$ and $\tilde{\bf Y}$ involves the hermitian
matrix product of two component spinors for which we have the following
identity:
\begin{equation}\acsz\begin{array}{rl}
\widetilde{
(\Phi_\sigma{\Psi_\sigma}^\dagger - \Psi_\sigma{\Phi_\sigma}^\dagger)}
={}& (\Phi_\sigma{\Psi_\sigma}^\dagger - \Psi_\sigma{\Phi_\sigma}^\dagger)
- \left(\tr{\Phi_\sigma{\Psi_\sigma}^\dagger -
\Psi_\sigma{\Phi_\sigma}^\dagger}
\right)\\ \noalign{\medskip}
={}& (\Phi_\sigma{\Psi_\sigma}^\dagger - \Psi_\sigma{\Phi_\sigma}^\dagger)
+ ({\Psi_\sigma}^\dagger\Phi_\sigma - {\Phi_\sigma}^\dagger\Psi_\sigma)\,\one,
\end{array}
\end{equation}
where $\sigma \in \{+,-\}$.
This relationship allows us to rewrite (\ref{eq:Fierz1}):
\begin{equation}\acsz\begin{array}{rl}
{\bf Y} ={}&(\Phi_+ {\Psi_+}^\dagger - \Psi_+ {\Phi_+}^\dagger)
+ (\Phi_-{\Psi_-}^\dagger - \Psi_-{\Phi_-}^\dagger)
+ ({\Psi_-}^\dagger\Phi_- - {\Phi_-}^\dagger\Psi_-)\,\one,
\\ \noalign{\medskip}
\tilde{\bf Y} ={}&(\Phi_+ {\Psi_+}^\dagger - \Psi_+ {\Phi_+}^\dagger)
+({\Psi_+}^\dagger\Phi_+ - {\Phi_+}^\dagger\Psi_+)\,\one
+ (\Phi_-{\Psi_-}^\dagger - \Psi_-{\Phi_-}^\dagger).
\end{array}
\label{eq:Fierz}
\end{equation}
These identities are plausible because of equation (\ref{eq:ymu}).
To prove them we need to use the fact that the $\Gamma_\mu$ are a basis for
the space of the hermitian matrices.  (Note that the matrices are grouped so
that the combinations in parentheses are hermitian.  In particular, their
traces are real, which means that we may commute octonionic products and/or
take octonionic conjugates.)

The distinct advantage of the octonionic formalism is that we can deal with
spinors without having to use Dirac matrices.  In fact, in the rest of the
article Dirac matrices only appear to relate our results to the usual notation,
but they are not used in any of the derivations.

\subsection{Lorentz transformations}\label{subsec:Lorentz}

In Clifford language the orthogonal group is the Clifford group which is
generated by unit vectors. A unit vector $p$ induces a reflection at a
line both on vectors and on spinors via the transformations
\begin{equation}\begin{array}{rcl}
\Slash{x}&\rightarrow&\Slash{p}\Slash{x}{\Slash{p}}^{-1},\\
\noalign{\medskip}
\Psi&\rightarrow&\Slash{p}\Psi.
\end{array}\label{reflection}\end{equation}
Components of vectors parallel to $p$ remain fixed, whereas those
perpendicular to $p$ are inverted.  A pair of unit vectors $p,q$ induces a
rotation in the plane spanned by them, which means that the even part of the
Clifford group corresponds to the simple orthogonal group.  Specifically the
9+1-dimensional proper orthochronous Lorentz transformations are generated by
\begin{equation}\begin{array}{rcl}
{\bf X}&\rightarrow&{\bf P}(\tilde{\bf Q}{\bf X}\tilde{\bf Q}){\bf P},\\
\noalign{\medskip}
\Psi_+&\rightarrow&{\bf P}(\tilde{\bf Q}\Psi_+),\\ \noalign{\medskip}
\Psi_-&\rightarrow&\tilde{\bf P}({\bf Q}\Psi_-),
\end{array}\label{eq:proper Lorentz}\end{equation}
where $p_\mu p^\mu = q_\mu q^\mu = 1$.
More details specifically about the effects of the non-associativity of the
octonions are given in \cite{Clifford,Lorentz}.

\subsection{The 3-$\Psi$'s rule}

The previous relationships (\ref{eq:Fierz}), which represent part of the
octonionic analogue of the Fierz identities, allow us to deduce the 3-$\Psi$'s
rule for anticommuting 9+1-D Majorana-Weyl spinors:  (Given $\sigma \in
\{+,-\}$, we take $\Psi_k=P_\sigma\Psi_k$ $\;\;\forall\, k\in\{1,2,3\}$.)
\begin{equation}
\acsz\begin{array}{rl}
\gamma^\mu\Psi_1\adj{\Psi_2}\gamma_\mu\Psi_3
={}& \widetilde{(\Psi_2 {\Psi_3}^\dagger - \Psi_3 {\Psi_2}^\dagger)}\Psi_1
\\ \noalign{\smallskip}
={}& (\Psi_2 {\Psi_3}^\dagger - \Psi_3 {\Psi_2}^\dagger)\Psi_1
- \left(\Re\tr{\Psi_2 {\Psi_3}^\dagger - \Psi_3 {\Psi_2}^\dagger}\right)\Psi_1
\\ \noalign{\smallskip}
={}& (\Psi_2 {\Psi_3}^\dagger)\Psi_1 - (\Psi_3 {\Psi_2}^\dagger)\Psi_1
+ \Psi_1({\Psi_3}^\dagger \Psi_2) - \Psi_1({\Psi_2}^\dagger \Psi_3).
\end{array}
\end{equation}
When we add the terms generated by cyclic permutations of the spinors, we may
express the result in terms of associators of octonions.  We may even treat
both spinor components simultaneously by defining an associator for spinors:
\begin{equation}
[\Psi_1,{\Psi_2}^\dagger,\Psi_3] := \Psi_1({\Psi_2}^\dagger \Psi_3) - (\Psi_1
{\Psi_2}^\dagger)\Psi_3
\end{equation}
This spinor associator is just a shorthand for the following expression
involving associators of the components:
\begin{equation}
[\Psi_1,{\Psi_2}^\dagger,\Psi_3] =
\left (\begin{array}{c}
[\Psi_{1\,1},\Bar{\Psi_{2\,1}},\Psi_{3\,1}]
+ [\Psi_{1\,1},\Bar{\Psi_{2\,2}},\Psi_{3\,2}]\\[1ex] \relax%
[\Psi_{1\,2},\Bar{\Psi_{2\,1}},\Psi_{3\,1}]
+ [\Psi_{1\,2},\Bar{\Psi_{2\,2}},\Psi_{3\,2}]
\end{array}\right),\label{eq:spinor associator}\end{equation}
where $\Psi_\alpha = \left( {\Psi_{\alpha\,1} \atop \Psi_{\alpha\,2}}
\right)\quad (\alpha = 1,2,3)$.
The expression (\ref{eq:spinor associator}) is symmetric in the last two
anticommuting spinors $\Psi_2$ and $\Psi_3$:
\begin{equation}
[\Psi_1,{\Psi_2}^\dagger,\Psi_3] = [\Psi_1,{\Psi_3}^\dagger,\Psi_2].
\end{equation}
(The derivation is a little tricky.  It uses the facts that octonionic
conjugation of one of the arguments of an associator merely changes its sign,
that the associator for (non-Grassmann) octonions is an antisymmetric function
of its three arguments, and that the components of $\Psi_2$ and $\Psi_3$
appear symmetrically in (\ref{eq:spinor associator}).)
Therefore we see that
\begin{equation}
\acsz\begin{array}{rl}
\gamma^\mu\Psi_1\adj{\Psi_2}\gamma_\mu\Psi_3 + \text{\ cyclic}
={}&-[\Psi_2,{\Psi_3}^\dagger,\Psi_1] + [\Psi_3,{\Psi_2}^\dagger,\Psi_1]
+ [\Psi_1,{\Psi_3}^\dagger,\Psi_2] \\ \noalign{\medskip}
&{}-[\Psi_1,{\Psi_2}^\dagger,\Psi_3] + [\Psi_2,{\Psi_1}^\dagger,\Psi_3] -
[\Psi_3,{\Psi_1}^\dagger,\Psi_2] \\ \noalign{\medskip}
={}& 0.
\end{array}\label{eq:3psis}
\end{equation}
This identity is required for the Green-Schwarz superstring to exhibit global
fermionic supersymmetry \cite{Green-Schwarz}.  This derivation shows that the
3-$\Psi$'s rule is a direct consequence of the alternativity of the octonionic
algebra, i.e., the relevant part of the Fierz identities are naturally built
into the algebraic structure of the octonions.

\subsection{A note on dotted and undotted spinors}

In 3+1 dimensions, the usual notion of dotted and undotted spinors for a
complex representation of $\Cl(3,1)$ arises from two facts: complex
conjugation of the Dirac matrices induces another faithful irreducible
representation of the Clifford algebra $\Cl(3,1)$, and matrix transposition
induces a faithful representation of the opposite Clifford algebra
$\Cl_\opp(3,1)$.  $\Cl_\opp(3,1)$ is the algebra obtained by defining $a
\vm_\opp b = b \vm a$, where $\vm$ (resp.\ $\vm_\opp$) denotes multiplication
in the abstract algebra (resp.\ its opposite).
Therefore, the two irreducible representations $\Gamma$ and $\tilde\Gamma$ of
the even subalgebra $\Cl_0(3,1)$, are essentially just complex conjugates of
each other, more precisely they are related by charge conjugation:
\begin{equation}\begin{array}{rc}
&\tilde{\bf X}^{\dot B B}
= \epsilon^{\dot B \dot A}\Bar{({\bf X}_{A\dot A})}\epsilon^{A B}\\[2ex]
\iff & \tilde {\bf X} =
\left(\begin{array}{cc} 0 & -1 \\ 1 & \phantom{-}0 \end{array}\right)
\Bar{\bf X}
\left(\begin{array}{cc} 0 & -1 \\ 1 & \phantom{-}0 \end{array}\right).
\end{array}\label{eq:tilde2}\end{equation}
(Note that for $x\in\C$, (\ref{eq:9-1rep}) and (\ref{eq:9-1even rep}) define a
representation of $\Cl(3,1)$.)
This relationship (\ref{eq:tilde2}) still holds in the octonionic case,
although octonionic conjugation does not result in another representation, nor
does matrix transposition give a representation for the opposite Clifford
algebra, because octonionic multiplication is not commutative:
\begin{eqnarray}
\Bar{(\Slash{p}\Slash{q})} & \neq & \Bar{\Slash{p}}\Bar{\Slash{q}},
\label{eq:no rep1}\\ \noalign{\smallskip}
\T{(\Slash{p}\Slash{q})} & \neq & \T{\Slash{q}}\T{\Slash{p}}.
\label{eq:no rep2}
\end{eqnarray}
As a consequence $\Phi$, defined by
\begin{equation}\begin{array}{rc}
&\Phi^{\dot B} := \epsilon^{\dot B \dot A}\Bar{(\Psi_A)}
= \epsilon^{\dot B \dot A}\Psi_{\dot A}\\[2ex]
\iff&\Phi := \left(
\begin{array}{cc} 0 & -1 \\ 1 & \phantom{-}0 \end{array}\right)
\Bar{\Psi_+},
\end{array}\label{eq:tilde3}\end{equation}
does not transform like a negative chirality spinor according to
(\ref{eq:proper Lorentz}).

Therefore the naive expectation that there is no essential difference the
analogous octonionic and complex representations of $\Cl(9,1)$ and $\Cl(3,1)$
turns out to be wrong.
For this reason we prefer to use the original relationship (\ref{eq:tilde}) as
a definition for $\tilde {\bf X}$ rather than (\ref{eq:tilde2}).  Remarkably,
we never have to use (\ref{eq:tilde2}) in any derivation, which
confirms that this relationship is not of primary importance.

Hermitian conjugation is still an antiautomorphism:
\begin{equation}
(\Slash{p}\Slash{q})^\dagger = \Slash{q}^\dagger\Slash{p}^\dagger.
\end{equation}
We already used this fact to obtain a Dirac hermitian form which defines the
spinor adjoint, see (\ref{eq:D-adjoint}).
So only two pairs of the four spinor spaces with lowered/raised,
undotted/dotted indices are in close correspondence.
This difference is related to the fact that spinors can be both Majorana and
Weyl in 9+1 dimensions.

Actually, it is still possible to restore relations (\ref{eq:no rep1}) and
(\ref{eq:no rep2}).  Namely, one has to switch to the opposite octonionic
algebra.  For example, the octonionic conjugate of an octonionic
representation is another representation when the original octonionic product
is replaced by its opposite.  This idea of utilizing various octonionic
multiplication rules, for example, the rule for the opposite octonionic
algebra, will be pursued further in \cite{Clifford}.

\section{The superparticle action, the equations of motion and their
solution}\label{sec:superparticle}

The action in the Lagrangian form, or second order action, for the
CBS-superparticle \cite{CBS} is given by
\begin{equation}
S = \int d\tau L(\tau),
\end{equation}
where
\begin{equation}\acsz\begin{array}{rl}
L ={}&{1\over 2}e\pi_\mu\pi^\mu\\
\noalign{\smallskip}
={}&{1\over 4}e\>\tr{{\bf\Pi}\tilde{\bf\Pi}},\\
\noalign{\medskip}
\pi_\mu={}&e^{-1}[\dot x_\mu + \sum_{A=1}^N \Re \adj{\dot\theta^A}
\tilde\Gamma_\mu \theta^A]
= e^{-1}[\dot x_\mu + \sum_{A=1}^N \Re \adj{\dot\theta^A} \gamma_\mu
\theta^A],\\
\noalign{\medskip}
{\bf\Pi}={}&e^{-1}[\dot{\bf X} + \sum_{A=1}^N
(\dot\theta^A\theta^{A\,\dagger}-\theta^A\dot\theta^{A\,\dagger})],
\end{array}\end{equation}
and the variables describing the superparticle are its spacetime position
$x_\mu$, a set of $N$ Majorana-Weyl spinors $\theta^A$, and $e$ is the einbein
or induced metric on the worldline.
The following equations of motion are obtained from varying the
action,
\begin{itemize}
\item{}with respect to $e$:
\begin{equation}\acsz\begin{array}{rl}
\pi_\mu\pi^\mu ={}&0\\
\noalign{\medskip}
\iff\quad
\tr{{\bf\Pi}\tilde{\bf\Pi}}={}&0;
\end{array}\label{eq:pi lightlike}\end{equation}
\item{}with respect to $x$:
\begin{equation}\acsz\begin{array}{rl}
\dot\pi_\mu ={}&0\\
\noalign{\medskip}
\iff\quad
\dot{{\bf\Pi}}={}&0;
\end{array}\label{eq:pi constant}\end{equation}
\item{}and with respect to $\theta^A$:
\begin{equation}\acsz\begin{array}{rl}
\Slash{\pi}\dot\theta^A ={}&\pi_\mu\tilde{\Gamma}^\mu\dot\theta^A= 0\\
\noalign{\medskip}
\iff\quad \tilde{\bf\Pi} \dot\theta^A={}&0.
\end{array}\label{eq:theta}\end{equation}
\end{itemize}

We solve the algebraic equations for ${\bf \Pi}$ and $\dot\theta^A$.
Equations (\ref{eq:pi constant}) and (\ref{eq:pi lightlike}) imply that $\pi$
is a constant lightlike vector.  Such vectors can be parametrized by 9 even
Grassmann parameters $\{\pi_1,\ldots,\pi_9\}$.
This parametrization is unique for the future or past
light cone in the regular case, i.e., if at least one of these components is
invertible and therefore has non-zero body.  In this case
$\sum_{i=1}^9\pi_i^2$ is invertible and has up to a sign a unique square root
$\pi_0$, whence $\pi^+$ or $\pi^-$ is invertible.

Otherwise, in the singular case when all components of $\pi$ have zero body,
there may not exist any $\pi_0$ to make $\pi_\mu$ lightlike, or there may be
multiple possibilities.  (For example, if the spatial components are all zero,
then $\pi_0$ may be any even Grassmann number which squares to zero.  These
difficulties arise, because $x\mapsto x^2$ is not injective in the
neighborhood of zero.)  We do not have a parametrization of this variation of
the trivial solution.

We parametrize ${\bf \Pi}$ by two real even Grassmann numbers $|a|, |b|$ and
an even Grassmann unit octonion $\unit{r}$, where $\unit{r}
\Bar{\unit{r}} = 1$.
\begin{equation}
{\bf \Pi}
= \twobytwo{{|a|}^2 & |a| |b| \unit{r} \\|a| |b| \Bar{\unit{r}} & {|b|}^2}.
\label{eq:pi solution}\end{equation}
This parametrization does not cover the case where one of $\pi^+$ or $\pi^-$
is neither invertible nor a square.
If $|a| = 0$ or $|b| = 0$, then $\unit{r}$ is undetermined.
We can solve (\ref{eq:theta}) for ${\bf \Pi}$, even in the pathological cases,
by letting
\begin{equation}
\dot\theta^A = {\bf \Pi}\zeta^A
= \left( \begin{array}{c} \pi^+ \\ \Bar{\pi} \end{array} \right) \zeta_1^A
+ \left( \begin{array}{c} \pi^{\phantom{-}} \\ \pi^- \end{array} \right)
\zeta_2^A,
\label{eq:gen solution}
\end{equation}
where $\zeta^A$ is an odd Grassmannian spinor.  This solution relies on the
weak form of associativity, the so-called alternativity, of the octonions,
which makes products which involve not more then two full octonions and their
octonionic conjugates associative.  If $\pi^+$ (resp.\ $\pi^-$) is invertible,
we may redefine $\zeta_1^A \to \zeta_1^A-{\pi\over\pi^+}\zeta_2^A$ (resp.\
$\zeta_2^A \to \zeta_2^A-{\Bar{\pi}\over\pi^-}\zeta_2^A$) to see that our
solution only depends on one arbitrary odd Grassmann octonion function.

If we can write ${\bf \Pi}$ as in (\ref{eq:pi solution}), then
\begin{equation}
\dot\theta^A =
\left( \begin{array}{c} |a| \\ |b| \Bar{\unit{r}}  \end{array} \right)
(|a| \zeta_1^A + |b|\unit{r}\zeta_2^A) = \Psi_0 \zeta_0^A,
\label{eq:theta solution}\end{equation}
where $\Psi_0 = {|a| \choose |b| \Bar{\unit{r}}}$
is a commuting spinor and $\zeta_0$ is an arbitrary odd Grassmann octonion
function.  So we have given the general classical solution for the CBS
superparticle (except for a parametrization of the lightlike vector in the
pathological cases).  This parametrization depends on two real even parameters
$|a|$ and $|b|$ and one even unit octonion $\unit{r}$, and a set of $N$ odd
octonion functions of $\tau$, $\zeta_0^A$ .

Our solution, involving the asymmetric $\Psi_0 = {|a| \choose |b|
\Bar{\unit{r}}}$, raises questions about the Lorentz invariance of the
parametrization.  In the work of others \cite{Oda}, ${\bf \Pi}$ is
parametrized in terms of the most general possible commuting spinor $\Psi = {a
\choose b}$:
\begin{equation}
{\bf \Pi} = \Psi\Psi^\dagger
= \twobytwo{{|a|}^2 & a \Bar{b} \\ b \Bar{a} & {|b|}^2 }.
\end{equation}
However, this parametrization introduces a redundancy of 7 extra parameters
that correspond to an octonionic unit sphere $S^7$, since only the combination
$\unit{r} = {a \Bar{b} \over |a| |b|}$ enters into the off diagonal elements
of ${\bf \Pi}$.  For this general $\Psi = {a \choose b}$, $\dot\theta^A = \Psi
\xi^A$, with $\xi^A$ an arbitrary odd Grassmann octonion, is not necessarily a
solution of (\ref{eq:theta}) because of the non-associativity of the
octonions.
By choosing one specific $\Psi_0 = {|a| \choose |b| \Bar{\unit{r}}}$ we
removed the redundancy and obtained the general solution $\dot\theta^A =
\Psi_0 \zeta_0^A$.  In this case all products are associative because they
involve only two independent octonionic directions, namely $\zeta_0^A$ and
$\unit{r}$.

A recent article by Cederwall \& Preitschopf \cite{S7} proposes to modify the
octonionic product in a way that keeps track of non-associativity.  We can
apply their ideas to obtain an alternate form of the solution which avoids the
specification of $\Psi_0$.
\begin{equation}
\dot\theta^A = \Psi \cpr_a \zeta^A_a\> \quad\text{\ or\ }\quad \dot\theta^A
= \Psi \cpr_b \zeta^A_b,
\label{eq:solution}
\end{equation}
where $x \cpr_a y := |a|^{-2}(x\Bar{a})(ay)$ and similar for $\cpr_b$.  The
new Grassmann functions $\zeta^A_a$ and $\zeta^A_b$ are related to $\zeta_0^A$
via ${a\over|a|}\zeta_0^A = \zeta^A_a$ and ${b\over|b|}\zeta_0^A = \zeta^A_b$.
%

In line with \cite{S7} the proper interpretation of ${\bf\Pi} =
\Psi\Psi^\dagger$ is to view it as an element of $\R \times \O P^1$, $\O P^1$
being the octonionic projective line.  The sixteen parameters of $\Psi$ are
collapsed, using the even Grassmannian generalization of the Hopf \cite{Hopf}
map: $\R \times S^{15} \isom \R \times S^{8} \times S^7$.  The singularities
for $|a|$ or $|b|$ not invertible are caused by the fact that the particular
coordinate maps cannot be extended to include both of these ``points''.

Lorentz invariance is broken by specifying a certain
multiplication rule for the octonionic product.  Using a modified product
adapted to the spinor components can be seen to restore the Lorentz invariance.

 From any of the forms for $\dot\theta^A$ we get $\theta^A$ by simply
integrating the arbitrary odd Grassmann octonion function, using the form of
(\ref{eq:gen solution}), for example,
\begin{equation}
\theta^A = {\bf \Pi}\Zeta^A + \theta^A_0.
\label{eq:solution theta}
\end{equation}
So given ${\bf \Pi}$, $\theta^A$ is parametrized by an arbitrary odd Grassmann
octonion function $\Zeta^A$ and a constant anticommuting spinor $\theta^A_0$.
${\bf X}$ may now be computed by a simple integration.

The choice of $\Zeta^A$ corresponds to the freedom of the local fermionic
supersymmetry discussed in the next section.  In fact, this supersymmetry can
be used to eliminate $\Zeta^A$.

\section{The Jordan matrix formalism}\label{sec:Jordan}

This section carries on the attempts of Foot \& Joshi \cite{Foot and Joshi}
and G\"ursey \cite{Gursey 5by5} to understand the symmetries of the
superparticle.  These include global supersymmetry, a local bosonic symmetry,
usually called the $\lambda$-transformation, and a local fermionic
supersymmetry, the $\kappa$-transformation.  We combine a fermionic spinor
variable $\beta$ and a bosonic vector ${\bf B}$ and scalar $b$ into one
superspace object, namely a $3\times 3$ Jordan matrix $\JO{B}$:
\begin{equation}
\JO{B}=\twobytwo{{\bf B} & \beta \\ \beta^\dagger & b}.
\end{equation}
($\beta$ corresponds to a positive chirality spinor.)  The Jordan product for
Jordan matrices with Grassmannian entries is taken to be defined as in
\cite{Foot and Joshi}, which is equivalent to taking the hermitian part of the
matrix product:
\begin{equation}
\JO{A} \jo \JO{B} :=
{1 \over 2} \left(\JO{A}\JO{B} + (\JO{A}\JO{B})^\dagger \right).
\end{equation}

The results of section \ref{subsec:Lorentz} can be applied to obtain a
generating set for all Lorentz transformations for a Jordan matrix:
\begin{equation}
\JO{A}\rightarrow\JO{M}\JO{A}\JO{M}^\dagger
\text{, where\ }
\JO{M} = \twobytwo{{\bf M} & 0 \\ 0 & 1},\
{\bf M} = {\bf P} \tilde \Gamma_1
\text{, and\ } p_\mu p^\mu =1.
\end{equation}
($\tilde{\bf Q}$ in (\ref{eq:proper Lorentz}) has been replaced by the
constant $\tilde \Gamma_1$, which is purely real and allows us to move the
parentheses.  This subset of transformations, of course, still generates all
of the Lorentz transformations.)

For the superparticle we consider
\begin{equation}
\JO{X} =
\twobytwo{{\bf X} & e^{1\over 2}\theta \\ e^{1\over 2}\theta^\dagger & e}
\end{equation}
as the fundamental superspace matrix.
(It is clear from the solution in the previous section that the various
fermionic variables decouple, which reflects a symmetry of the Lagrangian.  In
this section, we consider a single fermionic variable, i.e., $N=1$.)
The global supersymmetry transformation may then be written as
\begin{equation}\acsz\begin{array}{rl}
(1 + \delta_\epsilon)\JO{X} ={}& \Z_\epsilon \jo \JO{X}\\ \noalign{\medskip}
={}&\twobytwo{\one & 2 e^{-{1\over 2}}\epsilon \ \\ 0 & 1} \jo
\twobytwo{{\bf X} & e^{1\over 2}\theta \\ e^{1\over 2}\theta^\dagger & e}.
\\ \noalign{\medskip}
={}&\twobytwo{
{\bf X} + (\epsilon\theta^\dagger-\theta\epsilon^\dagger)
& e^{1\over 2}(\theta+\epsilon)\\
e^{1\over 2}(\theta^\dagger+\epsilon^\dagger) & e}.
\end{array}
\end{equation}
Note that we used the non-hermitian matrix $\Z_\epsilon$ for this
transformation, which avoids the extension to larger matrices in
\cite{Gursey 5by5}.

The $\lambda$-transformation has a simple structure as well:
\begin{equation}\acsz\begin{array}{rl}
(1 + \delta_\lambda)\JO{X}
={}& \JO{X} \jo {\Z_{\lambda\dot\theta}}^\dagger\\ \noalign{\medskip}
={}&\twobytwo{{\bf X} & e^{1\over 2}\theta \\ e^{1\over 2}\theta^\dagger & e}
\jo \twobytwo{\one & 0\\ 2 e^{-{1\over 2}}\lambda{\dot\theta}^\dagger & 1}
\\ \noalign{\medskip}
={}&\twobytwo{ {\bf X} +
\lambda(\theta{\dot\theta}^\dagger-\dot\theta\theta^\dagger) &
e^{1\over 2}(\theta+\lambda\dot\theta)\\
e^{1\over 2}(\theta^\dagger+\lambda{\dot\theta}^\dagger) & e}.
\end{array}
\end{equation}

We can also construct a superspace variable that contains
the conjugate momentum ${\bf \Pi}$ of ${\bf X}$:
\begin{equation}\acsz\begin{array}{rl}
\JO{P}^\prime:={}
&e^{-1}\dot{\JO{X}} \jo {\Z_{\theta}}^\dagger\\ \noalign{\medskip}
={}&e^{-1}\twobytwo{\dot{\bf X} & \dot{(e^{1\over 2}\theta)}\\
\dot{(e^{1\over 2}\theta)}^\dagger&\dot{e}} \jo
\twobytwo{\one & 0\\ 2e^{-{1\over 2}}\theta^\dagger&1}\\ \noalign{\medskip}
={}&\twobytwo{
e^{-1}[\dot{\bf X} + (\dot\theta\theta^\dagger-\theta{\dot\theta}^\dagger)] &
e^{-{1\over 2}}\dot\theta + {3\over 2}\dot{e}e^{-{3\over 2}} \theta\\
e^{-{1\over 2}}{\dot\theta}^\dagger
+ {3\over 2}\dot{e}e^{-{3\over 2}} \theta^\dagger
& \dot e}\\ \noalign{\medskip}
={}&\twobytwo{{\bf\Pi} &
e^{-{1\over 2}}\dot\theta + {3\over 2}\dot{e}e^{-{3\over 2}} \theta\\
e^{-{1\over 2}}{\dot\theta}^\dagger
+ {3\over 2}\dot{e}e^{-{3\over 2}} \theta^\dagger
& \dot e}.
\label{eq:p3by3}
\end{array}\end{equation}
However, we prefer to work better to postulate another superspace variable as
the ``conjugate'' to $\JO{X}$:
\begin{equation}
\JO{P}:=
\twobytwo{
{\bf\Pi} & e^{-{1\over 2}}\dot\theta\\
e^{-{1\over 2}}{\dot\theta}^\dagger& 0}.
\end{equation}
It can be used to give a pretty form for the solution of the equations of
motion:
\begin{equation}
\JO{P} = (\phi_a \phi_a^\dagger)_{\cpr_a} \quad \text{\ resp.\ } \quad
\JO{P} = (\phi_b \phi_b^\dagger)_{\cpr_b},
\label{eq:pretty}
\end{equation}
where
\begin{equation}
\phi_a =
\left(\begin{array}{c}
a \\ b \\ e^{-{1\over 2}}\Bar{\zeta_a}
\end{array}\right)
\quad \text{\ resp.\ } \quad
\phi_b =
\left(\begin{array}{c}
a \\ b \\ e^{-{1\over 2}}\Bar{\zeta_b}
\end{array}\right),
\label{eq:phi3}
\end{equation}
products are evaluated using the modified product, taking the hermitian part
is implied. This causes the (3,3) component to vanish.  This form
(\ref{eq:pretty}) exactly reproduces (\ref{eq:solution}).  It can be
interpreted to be a Grassmannian extension of the octonionic projective line,
which can also be defined as the matrices which are idempotent up to scale:
\begin{equation}
(\JO{P}\jo\JO{P})_{\cpr_a}=(\tr{\JO{P}})\JO{P}
=(\JO{P}\jo\JO{P})_{\cpr_b}.
\end{equation}

The $\kappa$-transformation can also be obtained using $\JO{P}$:
\begin{equation}
\delta_\kappa\JO{X} =
4 \twobytwo{\zero & \theta \\ 0 & e^{1\over 2}} \jo
\left(\JO{P} \jo \twobytwo{\zero & \kappa\\ \kappa^\dagger & 0} \right).
\end{equation}
Taking a closer look at this local fermionic symmetry, we realize that
\begin{equation}\begin{array}{c}
\delta_\kappa\theta = {\bf \Pi}\kappa,
\quad
\delta_\kappa{\bf X} = \dot{\theta} \delta_\kappa \theta^\dagger -
\delta_\kappa \theta \dot{\theta}^\dagger,
\quad
\delta_\kappa e
= 2 (\dot{\theta}^\dagger \kappa - \kappa^\dagger \dot{\theta})\\
\noalign{\medskip}
\implies \quad
\delta_\kappa {\bf \Pi}
= 2[\kappa (\tilde{\bf \Pi}\dot{\theta})^\dagger
- (\tilde{\bf \Pi}\dot{\theta}) \kappa^\dagger],
\end{array}\end{equation}
i.e., on shell $\delta_\kappa {\bf \Pi} = 0$ and $\delta_\kappa\theta$ has the
form of the general solution for $\dot{\theta}$.
(The form of the transformation simplifies due to our choice to include the
scale in the definition of ${\bf \Pi}$.)
We see that the $\kappa$-supersymmetry can be used to absorb the arbitrary odd
Grassmann octonion function in the solution (\ref{eq:solution theta}) for
$\theta$, so that just a constant spinor remains.
Therefore all solutions are generated by acting with $\kappa$-transformations
on solutions of the form
\begin{equation}\acsz\begin{array}{rl}
\theta ={}& \theta_0,\\ \noalign{\medskip}
\dot{\bf X} ={}& e{\bf\Pi} \\ \noalign{\smallskip}
\quad \implies \quad {\bf X} ={}& E\,{\bf\Pi} + {\bf X}_0,
\end{array}\end{equation}
where $\theta_0$ is a constant anticommuting spinor, ${\bf X}_0$ is a constant
vector, ${\bf\Pi}$ is a constant lightlike vector, and $E=\int e(\tau)\,d\tau$
is the arclength along the worldline.

The Freudenthal product for Jordan matrices is defined by
\begin{equation}
\JO{X} \fr \JO{Y} :=
\JO{X} \jo \JO{Y}
- {1 \over 2}\JO{X}(\tr{\JO{Y}}) - {1 \over 2}(\tr{\JO{X}})\JO{Y}
+ {1 \over 2}\left[ (\tr{\JO{X}})(\tr{\JO{Y}})-\tr{\JO{X}\jo\JO{Y}} \right]
\text{\large\bf 1}.
\end{equation}
This notion can be extended to Grassmannian Jordan matrices.  The Lagrangian
$L$ for the superparticle is then given by the following form which has the
same appearance as the $E_6$ invariant trilinear form on the non-Grassmannian
Jordan algebra:
\begin{equation}
L = -\>\tr{(\JO{P} \fr \JO{X}) \jo \JO{P}}.
\label{eq:Lagrangian}
\end{equation}
Due to the antisymmetry in (\ref{eq:Lagrangian}) with respect to the spinor
variables only the $(3,3)$ component of $\JO{X}$ contributes, i.e., $\JO{X}$
could be replaced by $\twobytwo{\zero&0\\0&e}$:
\begin{equation}
L = -\>\tr{(\JO{P} \fr \twobytwo{\zero&0\\0&e}) \jo \JO{P}}.
\end{equation}
Also only the upper $2\times 2$ matrix ${\bf\Pi}$ in $\JO{P}$ contributes to
the Lagrangian, so that $\JO{P}$ can be replaced by $\JO{P}'$ in either form.
Alternate versions of this formalism, where the einbein $e$ is substituted by
$1$ in the superspace variable $\JO{X}$, are possible, since the trilinear
form only contains certain combinations of variables, i.e., products of two
vectors and a scalar or of two spinors and a vector, both of which have units
of length squared.
%

\section{conclusion}

We have demonstrated the usefulness of the octonionic formalism in several
ways in this article.  We have solved the classical equations of motion for
the CBS superparticle.  The question of Lorentz covariance of the solution was
answered using a modified octonionic product.  The local fermionic
transformation was seen to relate solutions and to absorb the arbitrary odd
Grassmann octonion function in the solution for the fermionic variable.  We
have been able to express Lorentz and all known supersymmetry transformations
in terms of Jordan products involving Jordan matrices with Grassmannian
entries.  However, the exact form of the objects that should be used in these
expressions was unclear because of the cancellations due to the anticommuting
variables.  We believe that an extension to the Green-Schwarz superstring will
fix the form of the expressions, if such an extension is possible.  Another
interesting avenue is to explore further the symmetries of the theory in terms
of the Jordan matrices.
In \cite{S7} group manifolds are generalized to $S^7$ transformations.  By
taking a varying octonionic product into account, it may be possible to
generalize (super) Lie groups in a similar way.
An extension of the octonionic formalism off shell is needed to lead to a
quantization of the theory in this formalism, but it may be the key to unlock
the mysteries of the superstring.

\acknowledgments

I am grateful to my advisor Corinne Manogue for comments and discussions.  I
wish to thank the Bayley family for establishing the Bayley Graduate
Fellowship, which I had the honor to receive this year.  This work was partly
funded by the NSF grant \# PHY92-08494.
\\
\hbox to \hsize{\hfil\hfil\hfil\hfil\it soli deo gloria \hfil}


\begin{references}

\bibitem{supersymmetry and octonions}
T.~Kugo, P.~Townsend, {\it Supersymmetry and the division algebras},
Nucl.\ Phys.\ B {\bf 221}, 357 (1983)\SC
%
I.~Bengtsson, M.~Cederwall, {\it Particles, twistors and the division
algebras}, Nucl.~Phys.~B~{\bf 302}, 81 (1988)\SC
%
J.M.~Evans, {\it Supersymmetric Yang-Mills theories and division algebras},
Nucl.~Phys.~B~{\bf 298}, 92 (1988).

\bibitem{Gursey 5by5}
F.~G\"ursey, {\it Super Poincar\'e Groups and Division Algebras}, Modern
Physics~A~{\bf 2}, 967 (1987)\SC
%
F.~G\"ursey, {\it Supergroups in Critical Dimensions and Division Algebras},
Monographs on Fundamental Physics, {\bf Proceedings of Capri Symposia
1983-1987}, ed.\ Buccella-Franco, Lecture Notes Series No.\ 15, American
Institute of Physics, 1990, p.~529.

\bibitem{Chung}
K.W.~Chung, A.~Sudbery, {\it Octonions and the Lorentz and conformal groups
of ten-dimensional space-time}, Phys.~Lett.~B~{\bf 198}, 161 (1987)\SC

\bibitem{superstring and superparticle}
C.A.~Manogue, A.~Sudbery, {\it General solutions of covariant superstring
equations of motion}, Phys.~Rev.~D~{\bf 40} 4073 (1989)\SC
%
T.~Kimura, I.~Oda, {\it Superparticles and division algebras --- six
dimensions and quaternions}, Prog.~Theor.~Phys.~{\bf 80}, 1 (1988)\SC
%
I.~Bengtsson, {\it The fermionic gauge symmetry in the Green-Schwarz action},
Class.~Quantum Grav.~{\bf 4}, 1143 (1987)\SC
%
D.B.~Fairlie, C.A.~Manogue, {\it A parameterization of the covariant
superstring}, Phys.~Rev.~D~{\bf 36}, 475 (1987).

\bibitem{CBS}
R.~Casalbuoni, {\it The classical mechanics for Bose-Fermi systems}, Il~Nuovo
Cimento~A~{\bf 33}, 389 (1976)\SC
%
L.~Brink, J.H.~Schwarz, {\it Quantum superspace}, Physics~Letters~B~{\bf
100}, 310 (1981).

\bibitem{Oda}
I.~Oda, T.~Kimura, A.~Nakamura, {\it Superparticles and division algebras ---
ten dimensions and octonions}, Prog.\ Theor.\ Phys.~{\bf 80}, 367 (1988).

\bibitem{Tachibana}
H.~Tachibana, K.~Imaeda, {\it Octonions,
superstrings and ten-dimensional spinors}, Il~Nuovo Cimento~B~{\bf 104},
91 (1989).

\bibitem{Jordan}
E.~Corrigan,  T.J.~Hollowood, {\it The Exceptional Jordan Algebra and the
Superstring}, Commun.~Math.~Phys.~{\bf 122}, 393 (1989)\SC
%
E.~Corrigan,  T.J.~Hollowood, {\it A string construction of a commutative
non-associative algebra related to the exceptional Jordan algebra},
Physics~Letters~B~{\bf 203}, 47 (1988)\SC
%
R.~Foot, G.C.~Joshi, {\it String theories and Jordan algebras},
Physics~Letters~B~{\bf 199}, 203 (1987)\SC
%
F.~G\"ursey, {\it Discrete Jordan algebras and superstring symmetries}, {\bf
Proceedings of the John Hopkins workshop on current problems in particle
theory 13}, ed.\ L.~Lusanna,
World Scientific, Singapore 1989\SC
%
M.~G\"unaydin, G.~Sierra, P.K.~Townsend, {\it The geometry of $N=2$
Maxwell-Einstein supergravity and Jordan algebras}, Nucl.~Phys.~B~{\bf 242},
244 (1984)\SC
%
M.~G\"unaydin, G.~Sierra, P.K.~Townsend, {\it Exceptional supergravity
theories and the magic square}, Nucl.~Phys.~B~{\bf 242}, 244 (1984)\SC
%
G.~Sierra, {\it An application of the theories of Jordan algebras and
Freudenthal triple systems to particles and strings},
Class.~Quantum~Grav.~{\bf 4}, 227 (1987).

\bibitem{Foot and Joshi}
R.~Foot, G.C.~Joshi, {\it Space-time symmetries of superstring and Jordan
algebras}, Int.~J.~Theor.~Phys.~{\bf 28} 1449 (1989).

\bibitem{Fierz}
M.~Fierz, {\it Zur Fermischen Theorie des $\beta$-Zerfalls}, Zeitschrift f\"ur
Physik~{\bf 104}, 553 (1937).

\bibitem{others}
A.~Ferber, {\it Supertwistors and conformal supersymmetry}, Nucl.~Phys.~B~{\bf
132}, 55 (1978)\SC
%
W.~Siegel, {\it Hidden local supersymmetry in the supersymmetric particle
action}, Physics~Letters~B~{\bf 128}, 397 (1983),{\it Spacetime-supersymmetric
quantum mechanics}, Class.~Quantum Grav.~{\bf 2}, L95 (1985)\SC
%
D.P.~Sorokin, V.I.~Tkach, D.V.~Volkov, A.A.~Zheltukhin, {\it From the
superparticle Siegel symmetry to the spinning particle proper-time
supersymmetry}, Physics~Letters~B~{\bf 216}, 302 (1989)\SC
%
L.~Brink, S.~Deser, B.~Zumino, P.~Di~Vecchia, P.~Howe, {\it Local
supersymmetry for spinning particles}, Physics~Letters~B~{\bf 64}, 435
(1976)\SC
%
F.~Delduc, E.~Sokatchev, {\it Superparticle with extended worldline
supersymmetry}, Class.~Quantum Grav.~{\bf 9}, 361 (1992)\SC
%
A.S.~Galperin, P.S.~Howe, K.S.~Stelle, {\it The Superparticle and the
Lorentz group}, Nucl.~Phys.~B~{\bf 368}, 248 (1992)\SC
%
A.K.H.~Bengtsson, I.~Bengtsson, M.~Cederwall, N.~Linden, {\it Particles,
superparticles, and twistors}, Phys.~Rev.~D~{\bf 36}, 1766 (1987)\SC
%
M.~Cederwall, {\it A note on the relation between different forms of
superparticle dynamics}, hep-th/9310177.
%


\bibitem{Green-Schwarz}
M.B.~Green, J.H.~Schwarz, {\it Covariant description of superstrings},
Physics~Letters~B~{\bf 136}, 367 (1984).

\bibitem{Clifford}
J.~Schray, C.A.~Manogue, {\it Octonionic representations of
Clifford algebras and triality}, in preparation.

\bibitem{Lorentz}
C.A.~Manogue, J.~Schray, {\it Finite Lorentz transformations,
automorphisms, and division algebras}, J.~Math.~Phys.~{\bf 34}
3746 (1993).

\bibitem{octonions}
A.~Sudbery, {\it Division algebras, (pseudo) orthogonal groups and spinors},
Journal of Physics~A, mathematical and general {\bf 17} 939 (1984)\SC
%
R.D.~Schafer, {\bf An Introduction to Non-Associative Algebras} (Academic
Press, New York, 1966)\SC
%
M.~G\"unaydin, F.~G\"ursey, {\it Quark Structure and Octonions,}
J.~Math.~Phys.~{\bf 14}, 1651 (1973)\SC
%
A.~Gamba, {\it Peculiarities of the Eight-Dimensional Space},
J.~Math.~Phys.~{\bf 8}, 4 (1967).

\bibitem{general Clifford}
I.M.~Benn, R.W.~Tucker, {\bf An introduction to spinors and geometry with
applications  in physics}, (Adam Hilger, Bristol, Philadelphia, 1987)\SC
%
P.~Lounesto, {\it Scalar Products of Spinors and an Extension of Brauer-Wall
Groups}, Foundations of Physics {\bf 11}, 721 (1981)\SC
%
P.~Budinich, A.~Trautman, {\bf The Spinorial Chessboard} (Springer, Berlin
Heidelberg, 1988).

\bibitem{S7}
M.~Cederwall, C.R.~Preitschopf, {\it $S^7$ and $\widehat{S^7}$},
hep-th/9309030, (1993).

\bibitem{Hopf}
H.~Hopf, {\it \"Uber die Abbildungen der dreidimensionalen Sph\"are auf die
Kugelfl\"ache}, Mathematische Annalen~{\bf 104} 637 (1931);
\ifpreprintsty \\ H.~Hopf,
\fi\ {\it \"Uber die Abbildungen von Sph\"aren auf Sph\"aren niedriger
Dimension}, Fund. Math. {\bf 25}, 427 (1935).

\end{references}
\end{document}